\documentclass{edm_article}

\usepackage{graphicx}
\usepackage{url}
\usepackage{hyperref}
\usepackage{amsmath}
\usepackage{booktabs}
\usepackage{listings}
\usepackage{xcolor}

\lstdefinestyle{promptstyle}{
  basicstyle=\ttfamily\small,
  breaklines=true,
  breakatwhitespace=true,
  keepspaces=true,
  showstringspaces=false,
  frame=single,
  framerule=0pt,
  backgroundcolor=\color{gray!10},
  xleftmargin=0pt,
  xrightmargin=0pt,
  linewidth=\columnwidth,
  framexleftmargin=0.8em,
  framexrightmargin=0.8em,
  aboveskip=0.6\baselineskip,
  belowskip=0.6\baselineskip,
}

\lstdefinestyle{javastyle}{
  language=Java,
  basicstyle=\ttfamily\small,
  breaklines=true,
  breakatwhitespace=true,
  keepspaces=true,
  showstringspaces=false,
  frame=single,
  framerule=0pt,
  backgroundcolor=\color{gray!10},
  xleftmargin=0pt,
  xrightmargin=0pt,
  linewidth=\columnwidth,
  framexleftmargin=0.8em,
  framexrightmargin=0.8em,
  aboveskip=0.6\baselineskip,
  belowskip=0.6\baselineskip,
  keywordstyle=\color{blue}\bfseries,
  commentstyle=\color{gray}\itshape,
  stringstyle=\color{red},
}

\newcommand{\digitalappendix}{\url{https://github.com/ErrorSimulEDM/Digital-Appendix}}
\newcommand{\githublink}{\url{https://github.com/ErrorSimulEDM/Code}}

\begin{document}

\title{Simulating Students' Java Programming Errors with Large Language Models}
\numberofauthors{5}
\author{
\alignauthor
Ali Keramati$^*$\\
       \affaddr{University of California, Irvine}\\
       \email{a.kera@uci.edu}
\alignauthor
Jie Cao$^*$\\
       \affaddr{The University of North Carolina at Chapel Hill}\\
       \email{jiecao@unc.edu}
\alignauthor
Iman Mohammadi\\
       \affaddr{University of California, Irvine}\\
       \email{imohamm1@uci.edu}
\and
\alignauthor
Mark Warschauer\\
       \affaddr{University of California, Irvine}\\
       \email{markw@uci.edu}
\alignauthor
Yang Shi\\
       \affaddr{Utah State University}\\
       \email{yang.shi@usu.edu}
}

\maketitle
\footnotetext[1]{$^*$These authors contributed equally to this work.}

\begin{abstract}
Understanding student errors in the programming is a cornerstone of programming education, yet obtaining a representative set of student errors for any newly designed task remains slow and costly, since authentic submissions only accumulate after extensive classroom deployment. This paper explores whether large language models (LLMs) can serve as scalable proxies for students by simulating realistic logical errors in code submissions. Using the CodeWorkout dataset of 74,000+ unique student Java submissions across 37 problems, we evaluate five LLMs under three mainstream prompting strategies: Input-Output (IO), Chain-of-Thought (CoT), and iterative Self-Refine. We assess performance along two key dimensions: diversity (the range of distinct error patterns) and alignment (alignment with authentic student mistakes), and examine how these vary by struggling level of programming tasks. Our quantitative findings reveal that while all models generate diverse errors, their alignment to human submissions diverges: \textit{Claude Sonnet 4} achieves the most balanced performance. In addition, we conducted a blinded expert annotation study ($N=401$) comparing synthetic and authentic errors. This qualitative analysis confirms that the generated errors are functionally indistinguishable from authentic student errors. Moreover, higher-struggling-level problems elicit more diverse but less student-like errors. These results highlight trade-offs in using LLMs to simulate human learners and suggest design considerations for integrating synthetic errors into teachable agents, intelligent tutoring systems, and large-scale learning analytics.
\end{abstract}


\section{Introduction}
Programming education plays a foundational role in modern STEM curricula. Throughout this learning process, students inevitably make errors. Such errors not only expose students' weaknesses but also reveal their underlying cognitive processes—especially logical errors, which trace how learners reason about a problem.  Unlike syntactic mistakes, logical errors are more subtle: they reflect learners' incomplete understandings, misconceptions, and flawed problem-solving strategies. For decades, capturing such errors has been a cornerstone of learner modeling, personalized feedback, and adaptive instructional support in computer science education \cite{10.1145/2632320.2632343, 10.1145/3077618}. When teachers can foresee likely student errors, they can proactively design targeted scaffolds and address common pitfalls before they hinder student progress. Furthermore, integrating realistic error patterns into teachable agents can prompt learners to recognize and learn from these errors \cite{Matsuda2022}. Yet, anticipating the full spectrum of plausible student solutions and their corresponding failure modes for different tasks remains difficult, especially for novice educators or system designer.

Recent advances in Large Language Models (LLMs) suggest a promising path forward. LLMs can serve as scalable proxies for learners\cite{10874775, zhang-etal-2025-simulating,cao2026developing,HAO2026105472}, generating realistic error datasets that can power personalized feedback systems, intelligent tutoring environments, and deeper investigations into novice cognitive processes \cite{martynova-etal-2025-llms}. More broadly, LLMs are increasingly positioned as versatile tools for simulating students in learning analytics. This includes modeling learning behaviors \cite{2024.EDM-short-papers.31}, enriching classroom analytics with simulated reasoning \cite{zhang-etal-2025-simulating}, and reflecting the collaborative and motivational dynamics of online learning environments \cite{2025ORPs}.

Despite these advancements, the specific challenge of simulating \emph{logical} errors in programming education remains largely underexplored \cite{MacNeil_Synthetic}, and current approaches sit at two unsatisfying extremes. On one end, general prompting of LLMs typically yields correct or near-perfect solutions, entirely missing the characteristic flaws inherent to novice code \cite{wu-etal-2025-embracing, 10.1007/s10664-025-10614-4}. On the other, random corruption strategies (e.g., inserting arbitrary bugs) produce artificial mistakes that lack the pedagogical value of errors emerging from genuine, flawed reasoning. The challenge is thus one of indistinguishability a generated error must be wrong in ways a real student would be wrong---reflecting partial understandings, misconceptions, and incomplete problem-solving strategies rather than surface-level noise \cite{analytics2040046, 10.1109/ICSE55347.2025.00180}. Yet existing evaluation infrastructure offers little traction here. Programming benchmarks such as HumanEval, MBPP, and CodeContests are designed around absolute correctness, and recent work exploring synthetic buggy code has treated such code primarily as a vehicle for data augmentation \cite{10.1145/3716640.3716647} or focused on overall error distribution \cite{MacNeil_Synthetic} rather than logical errors

To evaluate whether LLMs can simulate students' logical errors, two properties must hold simultaneously: the generated errors need (i) span a sufficiently diverse space of failure modes, and (ii) exhibit high alignment by aligning with the actual errors students produce. Moreover, both properties may depend significantly on the intrinsic difficulty of the programming task. Problems that elicit numerous repeated attempts likely surface qualitatively different, and more varied, patterns of novice reasoning than those solved on the first try. We therefore operationalize the \textit{struggling level} of a problem using the total volume of student submissions as a behavioral proxy for empirical learning difficulty. Building on this perspective, we investigate how different combinations of models and prompting strategies influence the ability of LLMs to simulate student-like logical errors in programming. Specifically, we aim to answer the following research questions:

\begin{itemize}
    \item \textbf{RQ1 (Diversity):} How do different LLMs and prompting techniques vary in generating a diverse and expansive set of erroneous code?
    \item \textbf{RQ2 (Alignment):} To what extent does LLM-generated erroneous code align with real student logical errors?
    \item \textbf{RQ3 (Struggling Level as a Moderator):} How does the struggling level of problems affect LLMs' performance in simulating student-like logical errors?
\end{itemize}

\section{Related Work}
\subsection{Student Errors in Programming}
Errors are the most common byproduct of the students' learning process and serve as valuable learning resources. Teachers can analyze student errors to provide feedback and adjust their instructional plans, while students can engage in self-regulated learning through analyzing their own mistakes \cite{zimmerman2002becoming} or vicarious learning through erroneous examples \cite{Mayes2015}. In programming education, student errors have different types. Albrecht et al \cite{albrecht2020sometimes}, examined 12,371 submissions from 280 students and identified six error categories: syntactic, conceptual, strategic, sloppiness, misinterpretation, and domain knowledge errors.  According to other researchers,  logic errors represent the primary challenge that programming education must address \cite{qian2017students,macneil2024decoding}. Logic errors also encompass various subcategories. For instance, \cite{ettles2018common} analyzed 15,000 code fragments containing logic errors and classified them into algorithmic errors, misinterpretations of the problem, and fundamental misconceptions, finding that misconceptions are the most frequent source of logic errors.  Other researchers have identified additional logic error types, such as loop condition errors and logical operator misuse \cite{jackson2005identifying}.  By summarizing and analyzing these typical errors, valuable guidance can be provided for teachers’ instruction, or as the error proxy for students when they interact with teachable agent.

\subsection{LLM-Based Student Simulation}
Researchers have begun designing specific prompts to enable LLMs to assume particular roles or generate content that aligns with role-specific characteristics, leading to the emergence of student simulation research \cite{zhang-etal-2025-simulating,Pan2205,zhang2025seeking,cao2026developing}. These studies encompass three main approaches:
(1) direct agent simulation where LLMs embody virtual personas for flexible interactions, such as \cite{zheng2025cognitive} trained models to simulate students' cognitive abilities using transaction data from tutoring systems, and \cite{Pan2205} introduced TutorUp, a GPT-4-based platform enabling novice educators to practice engagement strategies through scenario-based interactions with simulated students;
(2) teacher-student dialogue simulation, exemplified by \cite{hu2025exploring}  utilized GPT-4 and Claude-3.5 to enhance teaching plan quality by simulating teacher-student interactions and generating teaching reflections;
(3) mimicking students' responses and behaviors, such as \cite{zhang2025seeking} build the students agent to respond to responsive teaching in teacher training. In the programming education domain, \cite{MacNeil_Synthetic} conducted a comparative study examining bug distributions generated by GPT-4 versus those produced by students, finding that while unguided LLMs do not generate plausible error distributions, they can be guided to produce realistic error patterns when provided with descriptions of common errors and typical frequencies. However, despite this preliminary exploration of LLM-based student programming error simulation, several gaps remain: the study lacks comparison across different LLMs, does not focus specifically on logical errors, and fails to examine the diversity of generated errors.

\section{Methodology}
\begin{figure}[h]
  \centering
  \Description{The process begins with (1) Data Preparation phase, where we scrape, merge, and filter the CodeWorkout dataset to create a clean set of Java programming problems. Next, in the (2) LLM Error Simulation phase, each problem is presented to a suite of LLMs using three distinct, parallel prompting strategies: an IO Prompting, a CoT Prompting, and an iterative Self-Refine loop}
  \includegraphics[width=\linewidth]{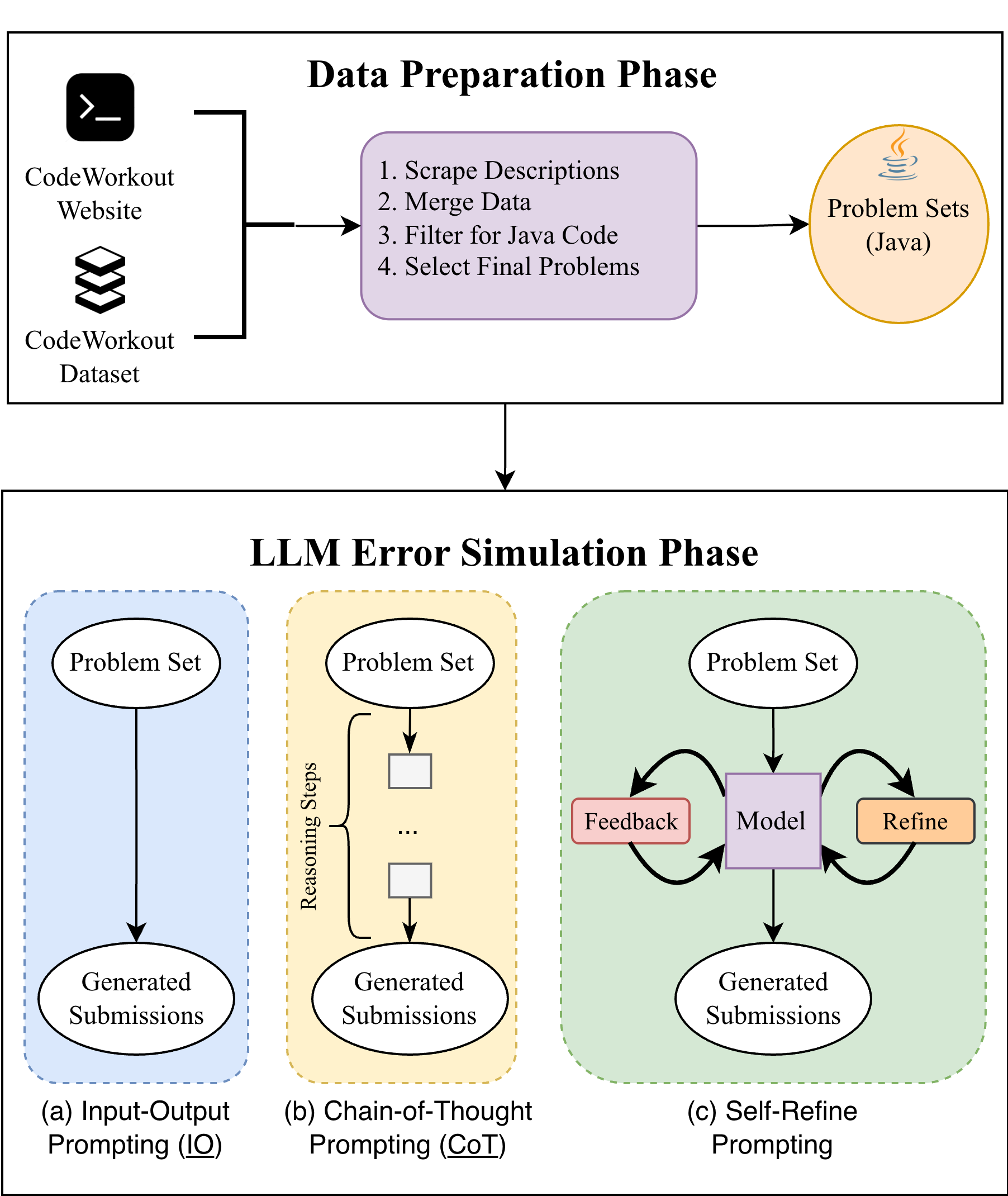}
  \caption{An overview of the experimental pipeline.}
  \label{figmet}
\end{figure}

To investigate to what extent LLMs can simulate logical errors in programming, we designed and executed a multi-stage methodology, leveraging the CodeWorkout dataset and a suite of prompting strategies across several LLMs, which is summarized in Figure~\ref{figmet}. The process begins with (1) Data Preparation phase, where we scrape, merge, and filter the CodeWorkout dataset to create a clean set of Java programming problems. Next, in the (2) LLM Error Simulation phase, each problem is presented to a suite of LLMs using three distinct, parallel prompting strategies: an IO Prompting, a CoT Prompting, and an iterative Self-Refine loop.

\subsection{Dataset}
We constructed our dataset from \textit{CodeWorkout}, which provides detailed logs of students' programming submissions for introductory Java exercises. These data were collected from 410 undergraduate students in an introductory Java course at Virginia Tech during the Spring and Fall 2019 semesters \cite{10.1145/3059009.3059055}. The raw data includes problem descriptions, code snapshots, and error-labeled submissions et al. To prepare it for analysis, we developed a preprocessing pipeline (Figure~\ref{figmet}) that transformed the raw logs into a clean, Java-focused corpus. Because large-scale CodeWorkout logs can contain abnormal or non-representative submissions (e.g., empty, duplicated, or otherwise anomalous traces), we follow prior work on detecting abnormal CodeWorkout submissions when considering data quality and filtering decisions \cite{hicks2024abnormal}. In addition, since the logs contained multiple programming languages, we applied a heuristic classifier based on regular expressions to identify Java submissions. Finally, to focus specifically on logical errors, we filtered out submissions that contained syntax errors (which fail to compile) and those that contained completely correct code. Consequently, the retained submissions consist exclusively of compilable yet incorrect code, ensuring the presence of logical errors. From the filtered corpus, we extracted the unique function names students were required to implement and aligned them with the scraped descriptions, ensuring all tasks used in our study were authentic CodeWorkout problems. The final dataset comprised 37 distinct Java problems. To capture variation in student struggle, we categorized them into three levels: Low (avg.\ 1{,}187 submissions), Medium (avg.\ 2{,}049) and High (avg.\ 4{,}125) based on the total number of student submissions per problem. Problems were sorted by submission count and evenly divided across levels (See detail in the \href{https://github.com/ErrorSimulEDM/Digital-Appendix}{Digital Appendix}\footnote{\url{https://github.com/ErrorSimulEDM/Digital-Appendix/blob/main/DigitalAppendix_ErrorSimul.pdf}} D, Table 5).

\subsection{Prompting Strategies}
\label{main:prompt_s}
We compared three prompting strategies: Input-Output (IO), Chain-of-Thought (CoT), and iterative Self-Refine. \textit{Input--Output prompting} is the standard direct-instruction baseline in which the model is given the task description and produces the final output in a single step, without articulating intermediate reasoning; it serves as the reference against which more elaborate strategies are compared \cite{10.5555/3600270.3602070}. \textit{Chain-of-Thought prompting} \cite{10.5555/3600270.3602070} instructs the model to articulate intermediate reasoning before emitting the final answer; in our setting, the model first writes a one- to three-sentence rationale describing the intended logical slip and then produces the buggy code. \textit{Self-Refine} \cite{10.5555/3666122.3668141} is an iterative framework in which the model generates an initial output, a critic instance produces structured feedback against a rubric, and a refiner instance rewrites items flagged for revision; this loop repeats for up to four iterations or terminates early when the critic judges the entire set compliant.
Each strategy was implemented as a distinct prompting pathway in our pipeline (Figure~\ref{figmet}), but all share the same experimental goal: generate multiple compilable Java submissions per problem, each containing exactly one non-trivial logical error and reflecting a plausible student attempt. The exact strategy instantiations, related prompts, and examples are documented in \href{https://github.com/ErrorSimulEDM/Digital-Appendix}{Digital Appendix A}.

\section{Experiments}
\subsection{Experimental Setup}
Our experiment was designed to compare the performance of different LLMs and prompting strategies in generating plausible, student-like programming errors for the 37 problems in our finalized dataset. As noted in the ``LLM Error Simulation'' phase of Figure~\ref{figmet}, we employed a diverse set of LLMs, including \texttt{google/gemini-2.5-pro}, \texttt{openai/gpt-5}, \texttt{openai/gpt-4o}, \texttt{anthropic/claude-sonnet-4}, and \texttt{x-ai/grok-code-fast-1}. These models were selected based on their performance on programming and code generation tasks at the time of this study, as indicated by leading industry benchmarks including the Vellum AI LLM Leaderboard \footnote{\url{https://www.vellum.ai/llm-leaderboard}}, LiveBench \footnote{\url{https://livebench.ai/}}, and OpenRouter \footnote{\url{https://openrouter.ai/rankings}}. All code generation tasks were performed with a sampling temperature of 0.0 to minimize output variance. An exception to this temperature setting was the feedback step in the Self-Refine strategy, which used a temperature of 0.3 to encourage more descriptive critiques. 

\subsection{Data Analysis}
\subsubsection{Automated Metrics}
For RQ1, we constructed abstract syntax trees (ASTs) for the generated erroneous code and computed pairwise edit distances using the Zhang–Shasha (ZSS) algorithm for each experimental condition \cite{zhangshasha}. Specifically, for each problem, we calculated the mean edit distance across the erroneous code samples, and then averaged these values across all problems. Larger distances indicate greater diversity in the generated erroneous code, meaning that within a given programming task, the more dissimilar the erroneous code samples are, the less consistent their error patterns become, thereby reflecting higher diversity.

For RQ2, to evaluate the alignment between LLM-generated and human erroneous code (fidelity), we first deduplicated student submissions, yielding 74{,}080 unique codes across 37 programming problems. Each LLM-generated erroneous code was then compared against human submissions for the corresponding problems. Since exhaustive all-pairs comparison was computationally infeasible (on average 2{,}002 human codes per task), we employed a two-stage matching strategy. In Stage~1, we represented each code as a bag of tokens and used Jaccard similarity to retrieve the top-$K$ most similar human codes for each LLM-generated submission ($K=100$, i.e., $\text{Recall}@100$ retrieval). In Stage~2, we applied the Zhang--Shasha tree edit distance within this candidate set to identify the nearest human match for each LLM-generated code. We use the ZSS distance to the nearest human match as our 
structural alignment metric (lower values indicate closer alignment 
with authentic student errors).

For RQ3, to investigate the effect of the struggling level of a problem, we compared edit distances across three struggling levels to analyze how the struggling level of a problem moderates both the diversity of LLM-generated erroneous code and their alignment with authentic student erroneous code. 

\subsubsection{Qualitative Evaluation}
While automated metrics provide a scalable measure of structural similarity, they do not fully capture the pedagogical utility of the generated errors. To address this limitation and to support construct validity, we conducted a qualitative study designed to assess whether the errors generated by the LLM are indistinguishable from authentic student errors to an expert instructor.

\textbf{Study Design and Protocol.} We employed a stratified sampling strategy to construct a balanced dataset of $N=444$ code submissions across all $37$ problems \cite{Braun01012006,AHMED2025100198}. For each problem we sampled $12$ submissions ($6$ authentic from the CodeWorkout logs and $6$ synthetic generated by our best-performing model, \textit{Claude Sonnet 4} with Self-Refine), yielding $37 \times 12 = 444$ items. All submissions were stripped of source-identifying metadata before annotation.

\textbf{Pilot Study and Codebook Development.} Two authors first conducted a pilot study on a random subset of $N=43$ submissions, double-coding each item independently. Through an iterative collaborative process, the annotators reconciled disagreements and established a consensus codebook \cite{doi:10.1177/16094069231205789} covering (i) eight logic-error categories with operational definitions and disambiguation rules (e.g., distinguishing Boundary errors from Condition Logic errors), and (ii) anchor descriptions for each level of the five-point plausibility scale, and (iii) whether the code was written by a Human Student or an LLM. The eight categories used in the final analysis (Table~\ref{tab:error_types}) were: \textit{Condition Logic} (faulty boolean expressions in \texttt{if}/\texttt{while} conditions); \textit{Boundary} (off-by-one and edge-case handling); \textit{Concept} (misunderstanding of the problem requirements or intended algorithm); \textit{Method} (incorrect use of library or built-in methods); \textit{Infinite Loop} (non-terminating control flow); \textit{Scope} (variable-scope or lifetime errors); \textit{Type} (type-related faults, e.g., \texttt{==} vs \texttt{.equals}); and \textit{Other} (a residual bucket for logical errors that did not map cleanly to a single named category, including strategy- or algorithm-level faults where the chosen approach itself is wrong, and, for authentic submissions, multi-fault or structurally awkward but compilable attempts).

\textbf{Annotation Phase.} The remaining $N=401$ submissions ($205$ authentic, $196$ synthetic) were divided between the two annotators, who each single-coded their assigned items under the finalized codebook with submissions blinded to source. Cases that the assigned annotator found ambiguous were brought to joint review and resolved by discussion. We did not compute a formal inter-rater reliability statistic on the main set because items were single-coded; the pilot phase served as the calibration step, after which the codebook was held fixed. The annotation criteria were:

\textbf{(1). Source Attribution (Turing Test):} A binary judgment on whether the code was written by a Human Student or an LLM.
\textbf{(2). Plausibility Score:} A five-point Likert rating ($1=$ Highly Artificial, $5=$ Highly Authentic) assessing whether the error reflects a genuine novice misconception.
\textbf{(3). Logic Error Taxonomy:} Categorization of the primary functional fault into one of the eight categories above.

\section{Results}
\subsection{RQ1: Diversity}
\textbf{LLMs can generate diverse erroneous code, but vary on different models and prompt strategies.} Overall, the results in Table~\ref{tab:mean_edit_distance_italic}  indicate that LLMs (across five models) are capable of producing diverse erroneous code, as reflected in the non-trivial mean edit distances observed across all models and prompting techniques. This suggests that, when prompted for multiple outputs on the same programming task, the generated code samples are not identical but vary in structure and content, thereby enabling the study of model-driven error diversity. However, some differences emerged both across models and across prompting strategies. At the model level, Gemini 2.5 Pro and Claude Sonnet 4 exhibited the largest average edit distance diversity, consistently exceeding 60 under most prompting techniques, indicating that they tend to generate more heterogeneous erroneous solutions. In contrast, GPT-4o demonstrated the smallest diversity, with mean edit distances around 25–40 depending on the prompt, suggesting more conservative or repetitive error patterns. GPT-5 and Grok Code Fast 1 occupied an intermediate position, with mean edit distances in the 40–60 range. Prompting strategies also modulated the diversity of erroneous outputs, though the effect was less uniform across models. For instance, in Claude Sonnet 4, the Self-Refine prompt reduced the mean edit distance compared to IO and CoT, suggesting more constrained revisions. By contrast, for Gemini 2.5 Pro, the CoT prompting increased variability, yielding the highest distance (72.55) observed in the dataset. Meanwhile, in models such as GPT-4o and Grok Code Fast 1, IO and Self-Refine performed comparably, with CoT prompting producing slightly higher diversity in some cases.

\begin{table*}[t]
\centering
\caption{Mean Edit Distance by Prompting Method and Model (values are mean $\pm$ SE).}
\label{tab:mean_edit_distance_italic}
\footnotesize
\setlength{\tabcolsep}{5pt}
\begin{tabular*}{\textwidth}{@{\extracolsep{\fill}}lccccc}
\toprule
 & \multicolumn{5}{c}{\textbf{Models}} \\
\cmidrule(lr){2-6}
\textbf{Method} & \textit{Claude Sonnet 4} & \textit{GPT-4o} & \textit{GPT-5} & \textit{Gemini 2.5 Pro} & \textit{Grok Code Fast 1} \\
\midrule
IO          & \textbf{63.33}$\pm$1.56 & 25.26$\pm$1.17 & 52.54$\pm$1.37 & 63.03$\pm$2.16 & 43.42$\pm$1.47 \\
CoT         & 61.34$\pm$1.92 & \textbf{39.93}$\pm$1.96 & \textbf{60.28}$\pm$1.59 & \textbf{72.55}$\pm$2.61 & \textbf{54.94}$\pm$2.41 \\
Self-Refine & 46.89$\pm$2.02 & 25.97$\pm$1.33 & 59.97$\pm$2.10 & 59.28$\pm$1.83 & 44.43$\pm$2.18 \\
\bottomrule
\end{tabular*}
\end{table*}

\subsection{RQ2: Alignment}
\textbf{LLMs can simulate student erroneous code, but vary on different models and prompt strategies.} We evaluated the alignment between LLM-generated erroneous code and human submissions by computing the mean of the nearest edit distance between each simulated code and its most similar human counterpart. Smaller distances indicate greater similarity to authentic student errors.

\subsubsection{Structural Alignment (AST Analysis)}
As shown in Table~\ref{tab:nearest_edit_distance_lower_is_better}, the results show that LLMs are capable of generating erroneous code that resembles student submissions (the nearest edit distance is 16.35), but the alignment of this simulation depends on the choice of model and prompting strategy. The distances are never negligible, meaning the simulation is not perfect, yet some models produce errors strikingly close to those made by students. Claude Sonnet 4 provides the most consistent alignment: across all prompting strategies, its distances remain relatively low, with the best case (16.35 under Self-refine) being the closest match to real student code among all conditions. Grok Code Fast 1 also demonstrates strong potential, achieving the absolute minimum distance (16.22) with Self-Refine. However, its performance fluctuates more across prompts (rising above 30 under IO and CoT), showing that while Grok can mimic student errors well, this depends more on the prompting technique. By contrast, GPT-5 and Gemini 2.5 Pro exhibit much larger distances, often exceeding 80, suggesting that although they generate erroneous code, the errors diverge considerably from authentic student patterns. GPT-4o lies in between: sometimes close to students (20.41 with CoT) but inconsistent, particularly when Self-Refine is used. Prompting strategies influence these outcomes. CoT generally improves alignment, especially for Claude and GPT-4o. Self-Refine has mixed effects: it delivers Grok’s best result but worsens GPT-4o’s alignment.

\begin{table*}[t]
\centering
\caption{Mean Nearest Edit Distance by Method and Model (values are mean $\pm$ SE).}
\label{tab:nearest_edit_distance_lower_is_better}
\footnotesize
\setlength{\tabcolsep}{5pt}
\begin{tabular*}{\textwidth}{@{\extracolsep{\fill}}lccccc}
\toprule
 & \multicolumn{5}{c}{\textbf{Models}} \\
\cmidrule(lr){2-6}
\textbf{Method} & \textit{Claude Sonnet 4} & \textit{GPT-4o} & \textit{GPT-5} & \textit{Gemini 2.5 Pro} & \textit{Grok Code Fast 1} \\
\midrule
IO          & 22.48$\pm$1.80 & 34.30$\pm$2.70 & 93.75$\pm$2.46 & \textbf{56.66}$\pm$3.85 & 39.66$\pm$3.06 \\
CoT         & 24.94$\pm$2.22 & \textbf{20.41}$\pm$2.15 & 100.56$\pm$2.47 & 87.44$\pm$3.48 & 33.54$\pm$3.07 \\
Self-Refine & \textbf{16.35}$\pm$1.80 & 49.85$\pm$3.08 & \textbf{80.59}$\pm$3.47 & 82.01$\pm$3.16 & \textbf{16.22}$\pm$1.78 \\
\bottomrule
\end{tabular*}
\end{table*}

\subsubsection{Human Evaluation}
\textbf{(1). Source attribution.}
For source attribution, we report the deception rate: the proportion of LLM-generated submission that annotators misclassified as human-written. By construction, this metric is defined only on the $196$ synthetic items. Annotators misclassified $164/196$ ($\mathbf{83.7\%}$) of LLM-generated submissions as human-written, indicating a low level of distinguishability.

\textbf{(2). Plausibility score:}
Contrary to the expectation that synthetic errors might be noisier or less coherent, the LLM-generated submissions received a significantly higher mean plausibility score ($M=4.27, SD=1.07$) than authentic student errors ($M=3.78, SD=1.11$). A Mann-Whitney U test confirmed this difference is statistically significant ($p < 0.001$).

This counterintuitive pattern is explained by the differences in shape between the two error distributions (Table~\ref{tab:error_types}). LLM-generated submissions concentrate on clean, single-purpose logical bugs, Condition Logic ($46.4\%$) and Boundary ($21.9\%$) faults dominate, which annotators read as canonical, textbook-style misconceptions. Authentic student code, although pre-filtered for compilability, exhibits a broader and more diffuse distribution: a residual $28.3\%$ of submissions fell into the \textit{Other} category, which captured logical errors that did not map cleanly to a single named type, including strategy- or algorithm-level faults and multi-fault attempts where several slips co-occurred in the same submission. Because the plausibility scale rewarded errors that read as a single, identifiable misconception, the LLM's narrower, well-formed bug distribution scored higher than the more diffuse one in authentic submissions. In practice, the LLM behaves like an idealized novice, valuable for isolating individual misconceptions when training intelligent tutoring systems, but a less faithful proxy for the full messiness of authentic student work.

\begin{table}[t!]
\centering
\caption{Qualitative Evaluation Results: Plausibility and Deception ($N=401$)}
\label{tab:qualitative_results}
\begin{tabular}{lcccc}
\toprule
\textbf{Source} & \textbf{N} & \textbf{Mean Plaus.} & \textbf{Std. Dev} & \textbf{Deception Rate} \\
\midrule
Human& 205 & 3.78 & 1.11 & - \\
LLM& 196 & \textbf{4.27} & 1.07 & \textbf{83.7\%} \\
\bottomrule
\multicolumn{5}{l}{\small *Mann-Whitney U: $p < 0.001$ }
\end{tabular}
\end{table}

\begin{table*}[t]
\centering
\footnotesize
\caption{Distribution of Error Types: Human students vs. LLM-generated.}
\Description{Authentic submissions are spread more broadly across categories, with $28.3\%$ falling into the residual \textit{Other} bucket used by annotators for logical errors that did not map cleanly to a single named category. Synthetic submissions concentrate on clean, single-fault semantic logic errors.}
\label{tab:error_types}
\resizebox{\textwidth}{!}{%
\begin{tabular}{lccccccccc}
\toprule
\textbf{Source} & \textbf{Condition} & \textbf{Boundary} & \textbf{Concept} & \textbf{Method} & \textbf{Infinite} & \textbf{Scope} & \textbf{Type} & \textbf{Other} \\
\midrule
Human students & \textbf{24.9}\% & 12.2\% & 14.6\% & 9.3\% & 1.5\% & 5.4\% & 3.9\% & 28.3\% \\
LLM-generated & \textbf{46.4\%} & \textbf{21.9\%} & 16.8\% & 5.6\% & 0.5\% & 0.0\% & 0.5\% & 7.1\% \\
\bottomrule
\end{tabular}%
}
\end{table*}

\subsection{RQ3: Struggling Level as a Moderator}

\begin{figure*}[h]
  \centering
  \Description{Mean edit distances among LLM submissions increased substantially from low- to medium-struggling problems, and remained high at high-struggling problems for most models.}
  \includegraphics[width=\linewidth]{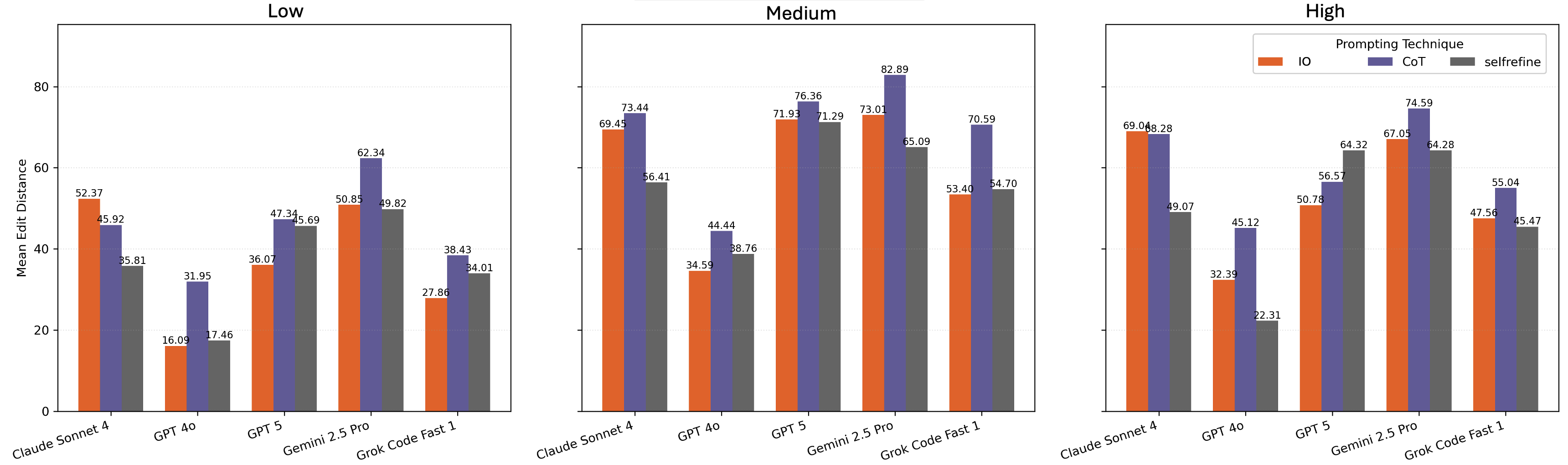}
  \caption{Mean Nearest Edit Distance Between LLM-Generated Errors Across Similar Tasks of Different Struggling Levels}
  \label{fig3}
\end{figure*}

\begin{figure}[h]
  \centering
  \Description{For problem ``sortasum'' (mean distance = 8.00), the model repeatedly produced nearly identical error patterns, with minimal structural variation. By contrast, in ``fix45'' (mean distance = 211.17), the generated errors were far more heterogeneous, with edit distances ranging from near zero to above 250.}
  \includegraphics[width=0.7\linewidth]{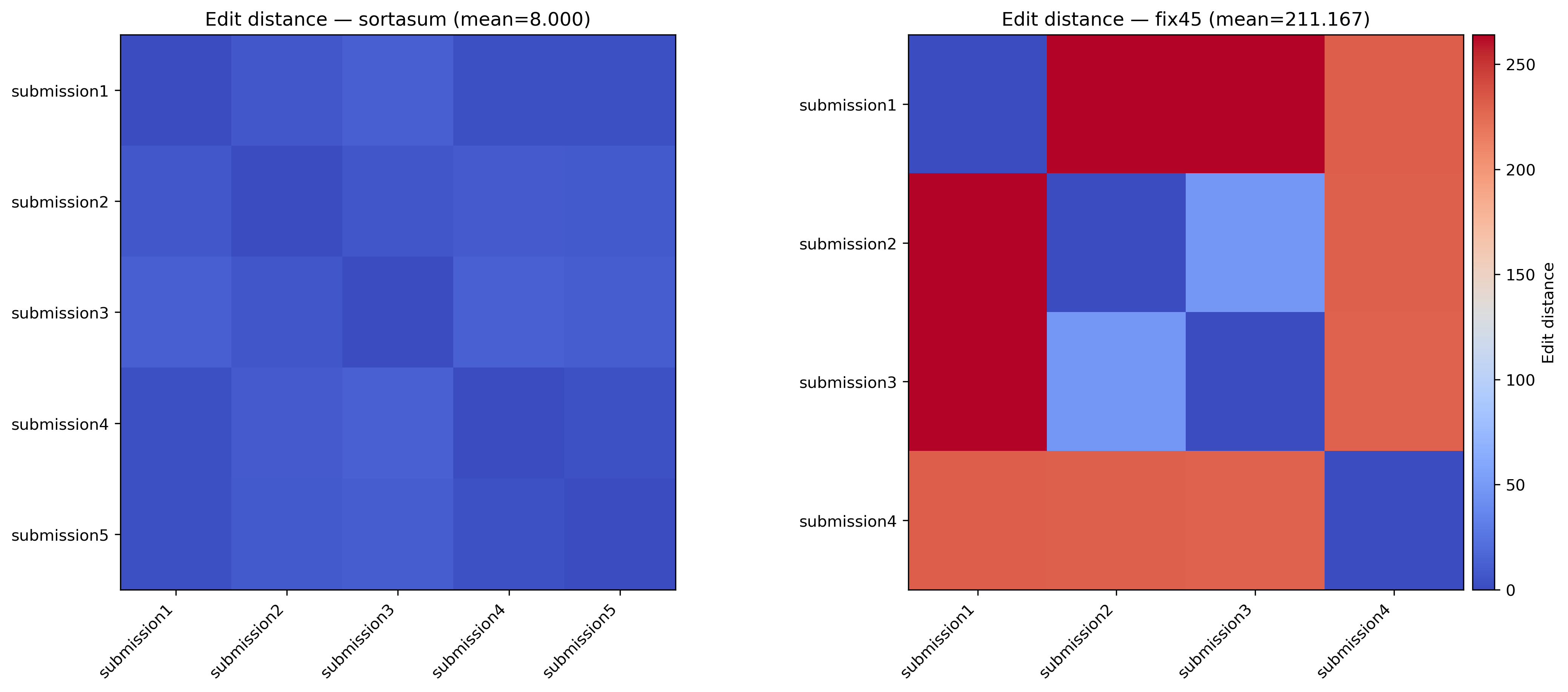}
  \caption{Pairwise Edit Distance Heatmaps Across LLM Code for Two Problems (Gemini 2.5 Pro, CoT Prompting).}
  \label{fig4}
\end{figure}

\textbf{LLMs simulate better in low-level questions but have higher diversity in high-level questions.}
As shown in Figure~\ref{fig3}, mean edit distances among LLM submissions increased substantially from low- to medium-struggling problems, and remained high at high-struggling problems for most models. This indicates that harder problems elicited more heterogeneous erroneous outputs, consistent with the expectation that higher-struggling problems open up more possible failure paths. For example, Gemini 2.5 Pro (CoT) rose from 62.34 (low) to 82.89 (medium) and stayed high at 74.59 (high). To probe the diversity of simulated errors, we conducted a case-level analysis using the best-performing setting (Gemini 2.5 Pro with CoT). For each coding problem, we measured the mean pairwise edit distance among all generated erroneous submissions. The results reveal strong problem-specific variation. For example, for problem ``sortasum'' (mean distance = 8.00), the model repeatedly produced nearly identical error patterns, with minimal structural variation. By contrast, in ``fix45'' (mean distance = 211.17), the generated errors were far more heterogeneous, with edit distances ranging from near zero to above 250. Heatmaps (Figure~\ref{fig4}) illustrate these two representative cases. Additional code examples for both problems are provided in \href{https://github.com/ErrorSimulEDM/Digital-Appendix}{Digital Appendix H}.

\begin{figure*}[h]
  \centering
  \Description{As the level of struggle increases, the mean nearest edit distance also increases, indicating that LLMs find it harder to reproduce student-like logical errors on higher-struggle problems}
  \includegraphics[width=\linewidth]{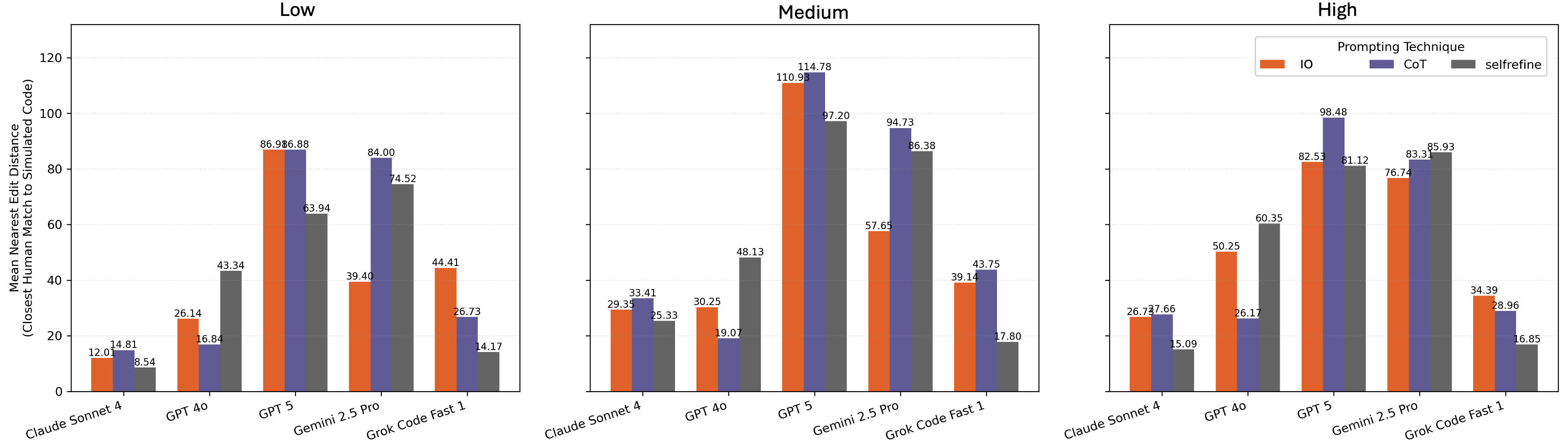}
  \caption{Mean Nearest Edit Distance Between LLM-Generated and Human Code Errors Across Tasks of Different Struggling Levels}
  \label{fig6}
\end{figure*}

\textbf{Alignment of LLM-generated code (simulated--human).} Figure~\ref{fig6} illustrates how the struggling level of coding problems affects the similarity between LLM-generated erroneous code and authentic student submissions. Overall, we observe a clear trend: as the level of struggle increases, the mean nearest edit distance also increases, indicating that LLMs find it harder to reproduce student-like logical errors on higher-struggle problems. Across models, Claude Sonnet 4 demonstrates the most stable performance, maintaining relatively low distances across all levels, but still has a slightly longer distance as the struggling level increases.  Moreover, other models, such as GPT-5 and Gemini 2.5 Pro, show substantial increases in distance at medium and high levels, indicating that their simulated errors diverge further from authentic student behavior as the struggling level rises. 

\section{Discussion and Conclusion}
\begin{figure}[h]
  \centering
  \Description{Models strong in diversity (e.g., Gemini) often exhibit weaker alignment, and vice versa, with Claude Sonnet 4 standing out as the most balanced option. }
  \includegraphics[width=\linewidth]{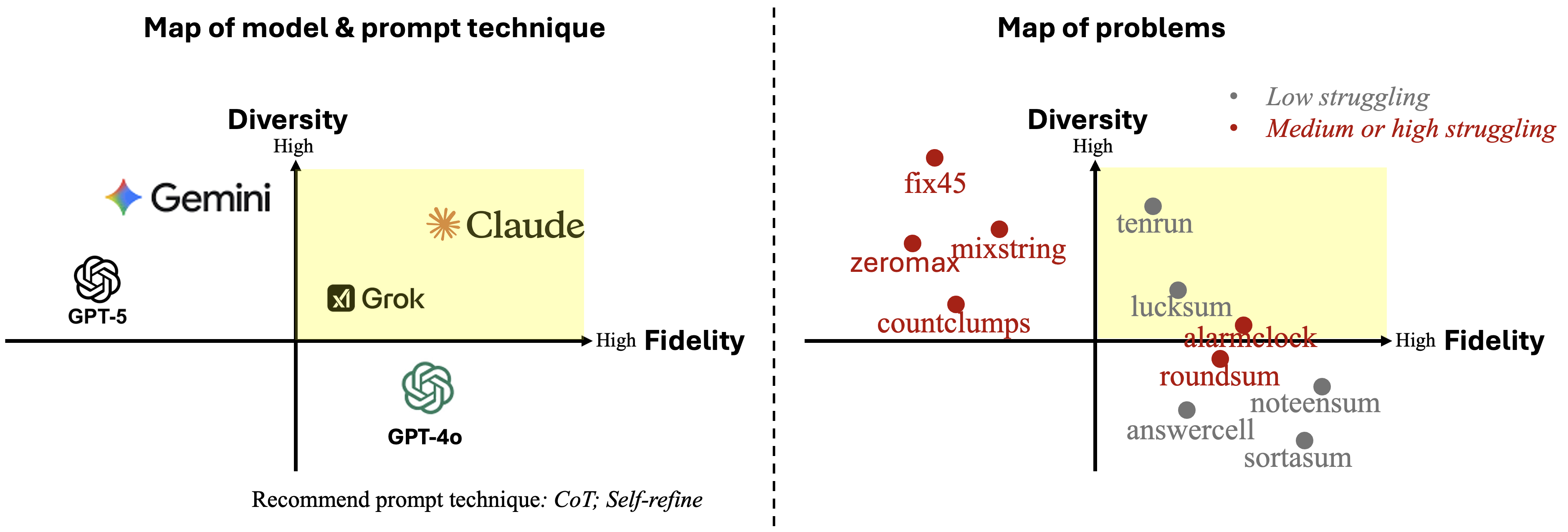}
  \caption{Trade-offs Between Alignment (fidelity) and Diversity in LLM-Simulated Student Errors}
  \label{fig7}
\end{figure}

This study investigated whether LLMs can simulate student-like logical errors in programming tasks and under what conditions. We identified two key dimensions, diversity and alignment, as central to their usefulness as proxies for authentic errors. On diversity, all models produced varied erroneous code, though to different extents: Gemini 2.5 Pro and Claude Sonnet 4 showed the broadest diversity (especially with CoT prompting), whereas GPT-4o generated more homogeneous patterns. On alignment (fidelity), i.e., resemblance to authentic erroneous code, Claude Sonnet 4, GPT-4o, and Grok Code Fast 1 aligned most closely with real submissions (notably with Self-Refine prompting), while GPT-5 and Gemini 2.5 Pro diverged more substantially. These trade-offs are illustrated in Figure~\ref{fig7}: models strong in diversity (e.g., Gemini) often exhibit weaker alignment, and vice versa, with Claude Sonnet 4 standing out as the most balanced option. Yet even Claude’s performance varied across tasks and prompts, underscoring that no single LLM is universally optimal. These findings suggest that model choice should be guided by analytic goals: diversity-oriented models can help explore the breadth of potential misconceptions, while alignment-oriented models are better suited for approximating authentic error distributions.

In addition, our results suggest that the coding problem-struggling level plays a dual role (also see Figure~\ref{fig7}): while a higher struggling level increases diversity of simulated errors, expanding the coverage of potential failure modes, it simultaneously reduces alignment between simulated and authentic code, suggesting that while models generate a wide variety of errors, these may diverge from the authentic patterns observed in learners. This duality highlights the importance of considering both diversity and fidelity when utilizing LLMs as proxies for student error data generation and analysis: models may excel at producing diverse erroneous code for higher-struggling tasks, but not all of those errors are truly representative of human misconceptions. Beyond the immediate evaluation of LLMs as error simulators, our findings intersect with ongoing discussions in learning analytics. In practical deployments, these diversity–fidelity trade-offs can be used to tune synthetic error generation toward either broad misconception coverage or closer approximation of observed classroom error distributions. The capacity of LLMs to generate diverse yet realistic errors offers a potential avenue for augmenting student data, particularly in contexts where authentic data is scarce, sensitive, or difficult to collect.

Finally, simulating students' programming errors using LLMs holds significant pedagogical value, particularly in addressing the challenge of data scarcity in educational technology. First and foremost, large-scale synthetic error datasets can be leveraged to train and evaluate automated tutoring systems (AI Tutors). This allows intelligent systems to better diagnose and provide feedback on diverse, realistic student misconceptions without being constrained by the limited availability of human-generated data. Second, from a student-centric perspective, these generated errors can be integrated into ``teachable agent'' environments. By tasking learners with diagnosing and correcting mistakes made by an AI proxy, the system fosters vicarious learning and deepens students' metacognitive debugging skills. Lastly, while it is impractical for educators to manually review thousands of synthetic code snippets, curated subsets of these LLM-generated errors can still serve as valuable resources for teacher professional development \cite{zhang2025seeking}. Exposure to a concentrated variety of anticipated misconceptions allows teachers to practice formulating effective pedagogical responses and design more targeted instructional strategies prior to actual instruction \cite{hu2025exploring}.

Despite these promising results, several limitations should be considered when interpreting our findings. Our experiments are grounded in CodeWorkout and a curated set of 37 introductory Java problems, so the observed diversity-alignment = trade-offs may not generalize to other institutions, curricula, assessment formats, or programming languages. We operationalized struggling level as the total number of submissions per problem, which is a scalable proxy but may conflate difficulty with other factors (e.g., assignment placement, popularity, or course policies); alternative measures, such as error rates, time-to-correct, or concept difficulty models, could yield different patterns. Furthermore, our generation constraints (compilable code with exactly one non-trivial logical error) intentionally reduce syntactic noise and multi-fault behaviors common in authentic novice submissions, potentially producing an ``idealized learner'' distribution that is useful for isolating misconceptions but less representative of messy real-world student work; future studies should extend evaluation across broader datasets, richer difficulty measures, and semantic metrics. Finally, while the AST edit distance serves as a robust quantitative proxy for measuring structural variation, we acknowledge its inherent limitations. First, raw numerical distances can lack direct interpretability context across different programming tasks. Second, as a purely structural metric, it fails to fully capture semantic diversity; functionally equivalent codes may exhibit distinct AST structures, and vice versa. Future research should consider employing normalized metrics alongside execution-based semantic evaluations (e.g., test-case execution traces) to provide a more holistic assessment of error diversity.

\clearpage

\section*{Acknowledgments}

This research was originally conducted during the LearnLab\textbf{ }Summer School in Carnegie Mellon University, with registration supported by CSSplice (NSF: \#2213789, \#2213790, \#2213791, \#2213792). This material is based in part upon work supported by the National Science Foundation under Grant No. 2315294. The authors gratefully acknowledge Utah State University for funding the API tokens.
\footnote{The source code associated with this study is available at: \githublink}
\footnote{Digital appendix is available at: \digitalappendix}

\bibliographystyle{abbrv}
\bibliography{sigproc}

\balancecolumns
\end{document}